\documentclass[12pt,a4wide]{article}

\usepackage{graphicx}
\usepackage{a4wide}
\usepackage{stmaryrd}
\usepackage{mathtools}
\usepackage{xcolor}
\usepackage{latexsym}
\usepackage{amssymb}
\usepackage{amsmath}
\usepackage{amsthm}
\usepackage{hyperref}
\usepackage{xspace}

\usepackage{multicol}

 \usepackage{lipsum}
 \usepackage{microtype}
 \usepackage{vwcol}  

\usepackage{bussproofs}
\usepackage{tikz}

\newtheorem{theorem}{Theorem}

\newtheorem{definition}[theorem]{Definition}
\newtheorem{example}[theorem]{Example}
\newtheorem{lemma}[theorem]{Lemma}
\newtheorem{proposition}[theorem]{Proposition}

\newcommand{\tempcut}[1]{{\color{brown} #1}}

\newcommand{\vcut}[1]{{\color{brown} #1}}

\renewcommand{\vcut}[1]{}
 \renewcommand{\tempcut}[1]{}

\newcommand{\parttwo}[1]{{\color{cyan} #1}} 
\renewcommand{\parttwo}[1]{}

\newcommand{\cut}[1]{}
\newcommand{\defstyle}[1]{\textbf{#1}}

\newcommand{\powerset}[1]{{\cal P}(#1)}

\newcommand{\frm}{\ensuremath{\mathcal{F}}}
\newcommand{\its}{\ensuremath{\mathcal{M}}}
\newcommand{\extn}[1]{[\![#1]\!]}

\newcommand{\pvar}{\ensuremath{\mathsf{VAR}}\xspace}
\newcommand{\rt}{\ensuremath{\mathbf{t}}\xspace}
\newcommand{\rn}{\mathord\ensuremath{\sim}\xspace}

\newcommand{\fu}{\ensuremath{\circ}\xspace}
\newcommand{\ri}{\ensuremath{\to}\xspace}

\newcommand{\RL}{\ensuremath{\mathcal{L}_R}\xspace}

\newcommand{\tr}{\ensuremath{\mathsf{Tr}}\xspace}

  \newcommand{\longver}[1]{#1}

 \newcommand{\shortver}[1]{}

\newcommand{\coimp}{\ensuremath{-\!\!\!\Yleft\,}\xspace} 
\newcommand{\intimp}{\ensuremath{\Rightarrow}\xspace} 
\newcommand{\furesfc}{\hookrightarrow} 

\newcommand{\complimp}{\ensuremath{\ri}\xspace}  
\newcommand{\complfu}{\ensuremath{\fu}\xspace}  
\newcommand{\complneg}{\ensuremath{\rn}\,\xspace}  

\newcommand{\qi}{\ensuremath{\implies}\xspace}

\newcommand{\sqema}{\ensuremath{\mathsf{SQEMA}}\xspace}
\newcommand{\alba}{\ensuremath{\mathsf{ALBA}}\xspace}
\newcommand{\algo}{\ensuremath{\mathsf{PEARL}}\xspace}

  \newcommand{\relml}[1]{}

\newcommand{\atoms}{\ensuremath{\mathsf{ATOMS}}\xspace}

\newcommand{\ineq}{\ensuremath{\leq}\xspace}

\newcommand{\ups}{\mathord{\uparrow}}
\newcommand{\dwns}{\mathord{\downarrow}}

\newcommand{\amp}{\mathop{\&}}

\newcommand{\nomi}{\mathbf{i}}
\newcommand{\nomj}{\mathbf{j}}
\newcommand{\nomk}{\mathbf{k}}
\newcommand{\cnomm}{\mathbf{m}}
\newcommand{\cnomn}{\mathbf{n}}

\newcommand{\RLP}{\ensuremath{\mathcal{L}^{+}_R}\xspace}

\newcommand{\nom}{\ensuremath{\mathsf{NOM}}\xspace}
\newcommand{\cnom}{\ensuremath{\mathsf{CNOM}}\xspace}

\newcommand{\FOR}{\ensuremath{\mathrm{FO}_{R}}}    

\begin{document}
\title{Algorithmic correspondence for relevance logics, bunched implication logics, and relation algebras:  \\ 
the algorithm \algo and its implementation (Technical Report)} 

\author{Willem Conradie$^{1}$
\and
Valentin Goranko$^{2}$
\and
Peter Jipsen$^{3}$
\\ ~ \\ 
$^{1}$School of Mathematics, University of the Witwatersrand, South Africa \\
Email: \textsf{willem.conradie@wits.ac.za}
\and
$^{2}$Department of Philosophy, Stockholm University, Sweden \\
Email: \textsf{valentin.goranko@philosophy.su.se} 
\and
$^{3}$Department of Mathematics, Chapman University, USA \\ 
Email: \textsf{jipsen@chapman.edu}
}

\maketitle

\begin{abstract}
\shortver{
The non-deterministic algorithmic procedure  \algo (acronym for `Propositional variables Elimination Algorithm for Relevance Logic') has been recently developed for computing first-order equivalents of formulas of the language of relevance logics \RL in terms of the standard Routley-Meyer relational semantics. It succeeds on a large class of axioms of relevance logics, including all so called inductive formulas. 
In the present work we re-interpret \algo from an algebraic perspective, with its rewrite rules seen as manipulating quasi-inequalities interpreted over Urquhart's relevant algebras, and report on its recent Python implementation.  We also show that all formulae on which \algo succeeds are canonical, i.e., preserved under canonical extensions of relevant algebras. This generalizes the ``canonicity via correspondence" result in \cite{DBLP:journals/sLogica/Urquhart96}. We also indicate that with minor modifications \algo can also be  applied to bunched implication algebras and relation algebras.
}

\longver{
The theory and methods of algorithmic correspondence theory for modal logics, developed over the past 20 years, have recently been extended to the language \RL of relevance logics with respect to their standard Routley-Meyer relational semantics. As a result, the non-deterministic algorithmic procedure \algo  (acronym for `Propositional variables Elimination Algorithm for Relevance Logic') has been developed for computing first-order equivalents of formulas of the language \RL in terms of that semantics. \algo  is an adaptation of the previously developed algorithmic procedures SQEMA (for normal modal logics) and ALBA (for distributive and non-distributive modal logics). It succeeds on all inductive formulas in the language \RL, in particular on all previously studied classes of Sahlqvist-van Benthem formulas for relevance logic.  

In the present work we re-interpret the algorithm \algo  from an algebraic perspective, with its rewrite rules seen as manipulating quasi-inequalities interpreted over Urquhart's relevant algebras. This enables us to complete the part of the Sahlqvist-van Benthem theorem still outstanding from the previous work, namely the fact that all inductive \RL-formulas are canonical, i.e., are preserved under canonical extensions of relevant algebras. Via the discrete duality between perfect relevant algebras and Routley-Meyer frames, this establishes the fact that all inductive \RL-formulas axiomatise logics which are complete with respect to first-order definable classes of Routley-Meyer frames. This generalizes the ``canonicity via correspondence" result in 
\cite{DBLP:journals/sLogica/Urquhart96} 
for (what we can now recognise as) a certain special subclass of Sahlqvist-van Benthem formulas in the ``groupoid'' sublanguage of \RL where fusion is the only connective. By extending \RL with a unary connective for converse and adding the necessary axioms, our results can also be applied to bunched implication algebras and relation algebras.

We then present an optimised and deterministic version of \algo , which we have recently implemented in Python and applied to verify the first-order equivalents of a number of important axioms  for relevance logics known from the literature, as well as on several new types of formulas. In the paper we report on the implementation and on some testing results.

}
\end{abstract}
%
%

\section{Introduction}
\label{sec:Introduction}

This work relates two important areas of development in non-classical logics, viz. \emph{relevance logics} and \emph{algorithmic correspondence theory}, by applying the latter to the possible worlds semantics for relevance logic based on Routley-Meyer frames \cite{Routleyetal82}, by means of an implementation of the recently developed in  \cite{ConradieGoranko2021} algorithm \algo.  That semantics is, in turn, duality-theoretically related to the algebraic semantics for relevance logic based on Urquhart's relevant algebras \cite{DBLP:journals/sLogica/Urquhart96}. Routley-Meyer frames also capture the semantics of (positive) relation algebras 
\cite{DBLP:journals/rsl/HirschM11}, \cite{DBLP:conf/concur/DoumaneP20-short}, and of bunched implication algebras \cite{pym2002semantics}, hence the algorithm \algo  implemented here is also applicable to arrow logic \cite{bent:note92}, \cite{DBLP:books/daglib/p/Dunn14} 
and bunched implication logics  \cite{pym2002semantics}. 

\paragraph{Modal Correspondence theory.}
 The Sahlqvist-van Benthem theorem \cite{Sahlqvist}, \cite{bent:moda76corr}, proved in the mid 1970s, is a fundamental result in the model theory of modal logic. 
  It gives a syntactic characterization of a class of modal formulas which define  first-order conditions on Kripke frames and which are canonical, hence, when added to the basic normal modal logic K, they axiomatize logics which are strongly complete with respect to elementary frame classes. The Sahlqvist-van Benthem theorem sets the stage for the emergence and development of the so called \emph{correspondence theory in modal logic}, cf. \cite{bent:corr2001}. The literature on the topic contains many analogues of the Sahlqvist-van Benthem theorem for a wide range of non-classical logics. Various illuminating alternative proofs have appeared, including Jonsson's purely algebraic proof of the canonicity part \cite{jonsson1994canonicity}, and the `canonicity-via-correspondence' approach pioneered by Sambin and Vaccaro \cite{sambin1989new}. 

The Sahlqvist-van Benthem class of formulas has been significantly extended to the class of so called \emph{inductive formulas}  \cite{Goranko:Vakarelov:2001,Goranko:Vakarelov:2002,GorankoVakarelovAPAL} 
which cover frame classes not definable by a Sahlqvist-van Benthem formula while enjoying the same properties of elementarity and canonicity. At about the same time, a new line of research known as \emph{algorithmic correspondence theory} emerged. It involves the use of algorithms\longver{\ for second-order quantifier elimination} like SCAN and DLS to try and compute first-order frame correspondence for modal formulas by eliminating the second-order quantifiers from their standard second-order frame correspondents. In particular, the algorithm \sqema  \cite{SQEMAI} was developed for algorithmic correspondence in modal logic. It  manipulates formulas in an extended hybrid language to eliminate propositional variables and thereby produces pure hybrid formulas which translate into first-order logic via the standard translation, and simultaneously proves their canonicity via an argument in the style of Sambin and Vaccaro. This approach was extended to  logics algebraically captured by normal (distributive) lattice expansions \cite{ALBAPaper,NonDistALBA} in a line of research that has become known as \emph{unified correspondence} \cite{UnifiedCor}.

\paragraph{Correspondence theory for Relevance logic.}
Much work has been done over the years on computing first-order equivalents and proving completeness of a range of specific axioms for relevance logics with respect to the \emph{Routley-Meyer relational semantics} (cf. \cite{Routleyetal82}). 
Routley-Meyer frames involve not a binary, but a ternary relation, with several conditions imposed on it, needed to ensure upward closedness of the valuations of all formulas. That makes the possible worlds semantics for relevance logic based on such frames  technically more complex and proving  correspondence  results for it ``by hand'' can be significantly more elaborate than those for modal logics with their standard Kripke semantics, which calls for  a systematic development of respective correspondence theory for relevance logics. 
Until recently, that problem remained little explored, with just a few works, incl. those of Seki \cite{seki2003sahlqvist} and Badia \cite{badia2018sahlqvist}, defining some classes of  Sahlqvist-van Benthem type formulas for relevance logics and proving correspondence results for them. 
Likewise, Suzuki \cite{suzuki_2011,suzuki_2013}, has established correspondence for the full Lambek calculus with respect to the so-called bi-approximation semantics, obtained via canonical extensions in the style of \cite{dunn2005canonical}.
For closely related distributive substructural logics, such as bunched implication logics, 
an elegant categorical approach to canonicity and correspondence is based on duality theory and coalgebras \cite{DahlPym2017}. 
\longver{The general utility of Sahlqvist-style results in this area of logic is witnessed by works like \cite{calcagno2007context} which proves completeness results for context logic and bunched logic via interpretation into modal logic and the application of the classical Sahlqvsit theorem. A similar approach is used in \cite{Brotherston:Calgano:ClassicalBI}  for classical bunched logic.  Lastly, \cite{Docherty:Pym:Dualities:Separation:Logic} develops Stone-type duality for a number of bunched logics including separation logic and positions this is the necessary preliminaries to a Sahlqvist theorem for these logics.}
\shortver{
A general algorithmic correspondence theory of relevance logics has recently been developed in \cite{ConradieGoranko2021}, on which the presently reported work is based.
}

 \longver{A general algorithmic correspondence theory of relevance logics has recently been developed in \cite{ConradieGoranko2021}, on which the presently reported work is based. That work includes the definition of the classes of inductive and Sahlqvist formulas for relevant logic. It  is shown  that \algo successfully computes first-order correspondents on Routley-Meyer frames for all such formulas. These definitions follow the general methodology of \cite{NonDistALBA} by which these classes are to be identified based on specific order-theoretic properties of the algebraic interpretations of the connectives. This gives a principled basis for comparing inductive and Sahlqvist formulas across different logics and different relational semantics for the same logic.}

\paragraph{The algorithm \algo and its implementation.}
A non-deterministic algorithmic procedure \algo (acronym for \textsf{Propositional variables Elimination Algorithm for Relevance Logic}) for computing first-order equivalents in terms of frame validity of formulas of the language \RL for relevance logics is developed in  \cite{ConradieGoranko2021}. 
 \algo is an adaptation of the above mentioned procedures \sqema \cite{SQEMAI} (for normal modal logics) and \alba  \cite{ALBAPaper,NonDistALBA} (for distributive and non-distributive modal logics).   
Furthermore, 
a large syntactically defined class of \emph{inductive relevance formulas} in \RL 
is defined in \cite{ConradieGoranko2021}, based on specific order-theoretic properties of the algebraic interpretations of the connectives, following the general methodology of \cite{NonDistALBA}. It is shown in \cite{ConradieGoranko2021}
that \algo succeeds for all such formulas and correctly computes for them equivalent with respect to frame validity first-order definable conditions on Routley-Meyer frames. This gives a general basis for comparing inductive and Sahlqvist formulas across different logics and for different relational semantics for the same logic. Thus, \cite[Example 3.14]{NonDistALBA} has shown that Suzuki's Sahlqvist class is properly included in the respective class of inductive formulas.  
Likewise, for the case of  \RL, it is shown in  \cite{ConradieGoranko2021} that the class of inductive formulas properly extends the classes of Sahlqvist formulas of Seki \cite{seki2003sahlqvist} and Badia \cite{badia2018sahlqvist}.

In the present work we re-interpret the algorithm \algo from an algebraic perspective with its rewrite rules seen as manipulating quasi-inequalities interpreted over Urquhart's relevant algebras 
\cite{DBLP:journals/sLogica/Urquhart96}. This enables us to complete the part of the Sahlqvist-van Benthem theorem still outstanding from the previous work, namely the fact that all inductive \RL-formulas are canonical, i.e., are preserved under canonical extensions of relevant algebras. Via the discrete duality between perfect relevant algebras and Routley-Meyer frames, this establishes the fact that all inductive \RL-formulas axiomatise logics which are complete with respect to first-order definable classes of Routley-Meyer frames. This generalizes the ``canonicity via correspondence" result in 
\cite{DBLP:journals/sLogica/Urquhart96}
for (what we can now recognise as) a certain special subclass of Sahlqvist-van Benthem formulas in the ``groupoid'' sublanguage of \RL where fusion is the only connective. 
We then present an optimised and deterministic version of \algo , which we have recently implemented in Python and applied to verify the first-order equivalents of a number of important axioms  for relevance logics known from the literature, as well as on several new types of formulas. In this paper we report on the implementation and on some testing results.

\paragraph{Relevance logics and relation algebras.} Even though developed with different motivations, these two areas are technically closely related, as noted and explored in several papers besides \cite{DBLP:journals/sLogica/Urquhart96}, incl. 
\cite{DBLP:journals/rsl/BimboDM09}, 
\cite{DBLP:journals/rsl/Maddux10}, 
\cite{DBLP:journals/rsl/HirschM11}, 
 \cite{TACL2013:Kowalski},  
 \cite{DBLP:books/daglib/p/Dunn14}.
 We note that, by extending \RL with a Heyting implication (which is a residual of the meet operation), removing relevant negation, and adding commutativity and associativity as axioms of fusion, our results can also be applied to bunched implication algebras. Alternatively one can extend \RL with classical implication and apply the same algorithm to relation algebras. In this case the Routley-Meyer frames have the order of an antichain and are the same as atom structures of relation algebras. Further details are discussed at the end of Section~\ref{sec:results}.

\paragraph{Structure of the paper.}
In Section \ref{sec:Preliminaries} we provide the necessary background on the syntax, algebraic and relational semantics of relevance logic, define relevant algebras and then extend their language by adding adjoints and residuals of the standard operators of relevance logic. 
Then, in  Section \ref{sec:Duality} we establish duality between perfect relevant algebras and complex algebras of Routley-Meyer frames. Section \ref{sec:Algo} presents the rules of the calculus on which \algo is based, and Section \ref{sec:Description} contains a concise description of the main phases of the algorithm itself. In Section \ref{sec:implementation} we give a brief description of the implementation of \algo, and in Section \ref{sec:results} we state some results. We then conclude with Section \ref{sec:Concluding}. After the references we have included a short appendix containing some additional technicalities and some examples of the output of \algo.  

\section{Preliminaries} 
\label{sec:Preliminaries}
In this section we provide background on the syntax and algebraic and relational semantics of relevance logic. For further details we refer the reader to \cite{Routleyetal82}, \cite{Dunn-Restall:02} and (for relevance logics) to \cite{DBLP:journals/sLogica/Urquhart96} and  \cite{ConradieGoranko2021}.

\subsection{Relevance logic and its algebraic semantics}
\label{subsec:RL+as}

The language of propositional relevance logic $\RL$ over a fixed set of propositional variables $\pvar$ is given by  
\[
A =p
\mid \bot \mid \top \mid \rt
\mid\, \rn A 
\mid (A \land A) 
\mid (A \lor A) 
\mid (A \fu A) 
\mid (A \ri A) 
\]
for $p \in \pvar$. The relevant connectives $\fu$, $\rn$ and $\ri$ are called \defstyle{fusion}, \defstyle{(relevant) negation} and \defstyle{(relevant) implication}, respectively. The constant $\rt$ is referred to as \defstyle{(relevant) truth}. We also add the constants $\top$ and $\bot$ for convenience. Equations and inequalities of $\RL$-formulas can be algebraically interpreted in relevant algebras as defined by Urquhart in \cite{DBLP:journals/sLogica/Urquhart96}.  

\begin{definition}[\cite{DBLP:journals/sLogica/Urquhart96}]\label{def:relevant:algebra}
	A structure $\mathbb{A} = \langle A, \wedge, \vee, \fu, \ri, \rn, \rt, \top, \bot \rangle$ is called a \defstyle{relevant algebra} if it satisfies the following conditions:
	\begin{enumerate}
	\begin{multicols}{2}
		\item $\langle A, \wedge, \vee, \top, \bot \rangle$ \\ 
		is a bounded distributive lattice,
		\item $a \fu (b \vee c) = (a \fu b) \vee (a \fu c)$,
		\item $(b \vee c) \fu a = (b \fu a) \vee (c \fu a)$,
		\item $\rn(a \vee b) = \rn a \wedge \rn b$,
		\item $\rn(a \wedge b) = \rn a \vee \rn b$,
		\item $\rn \top = \bot$ and $\rn \bot = \top$,
		\item $a \fu \bot = \bot \fu a = \bot$,
		\item $\rt \fu a = a$, and 
		\item $a \fu b \leq c$ iff $a \leq b \ri c$.
	\end{multicols}		
	\end{enumerate}
\end{definition}

An $\RL$-formula $\phi$ is \defstyle{valid on a relevant algebra $\mathbb{A}$} if the inequality $\rt \leq \phi$ (implicitly universally quantified over all propositional variables) is valid on $\mathbb{A}$ and \defstyle{valid on a class of relevant algebras} if it is valid on each member of that class. We also refer the reader to \cite{DBLP:journals/sLogica/Urquhart96} for axiomatizations of the logic of the class of all relevant algebras. %

\subsection{Relational semantics}

Relevance logic can be given relational semantics based on structures called `Routley-Meyer frames', which we will now define. A  \defstyle{relevance frame} is a tuple $\frm = \langle W, O, R, ^\ast  \rangle$, where: 

\begin{itemize}
	\item $W$ is a non-empty set of states (possible worlds); 
	\item $O \subseteq W$ is the subset of  \defstyle{normal} states;
	\item $R \subseteq W^{3}$ is a   \defstyle{relevant accessibility relation}; 
	\item $^\ast: W \to W$  is a function, called the \defstyle{Routley star}.  
\end{itemize}

The binary relation $\preceq$ is defined in every relevance frame by specifying that $u \preceq v \mbox{ iff } \exists o (o \in O \land R ouv)$. A  \defstyle{Routley-Meyer frame}
\footnote{
	The definition of Routley-Meyer frames takes the relation $R$ and subset $O$ as primary and defines the pre-order $\preceq$ in terms of them. This does not restrict the pre-orders that can occur within Routley-Meyer frames. 
	Indeed, given an upward closed subset $O \subseteq W$ and a pre-order $\preceq$ on $W$ one can define a respective ternary relation $R \subseteq W^3$ by specifying that, for all triples $(x,y,z)$, $Rxyz$ iff $x \preceq o$ for some $o \in O$ and $x \preceq y$. 
}
(for short, \defstyle{RM-frame}) is a relevance frame satisfying the following conditions for all $u,v,w,x,y,z \in W$: 
\begin{multicols}{2}
	\begin{enumerate}
		\item $x \preceq x$ 
		\item  If $x \preceq y$ and $Ryuv$ then $Rxuv$.
		\item  If $x \preceq y$ and  $Ruyv$ then  $Ruxv$.
		\item  If $x \preceq y$ and  $Ruvx$ then  $Ruvy$.
		\item  If $x \preceq y$ then $y^{\ast} \preceq x^{\ast}$.
		\item  $O$ is upward closed w.r.t. $\preceq$, \\ 
		i.e. if $o \in O$ and $o \preceq o'$ then $o' \in O$. 
	\end{enumerate}
\end{multicols}	

These properties ensure that $\preceq$ is {reflexive and} transitive, hence a preorder, and that the semantics of the logical connectives has the upward monotonicity property stated below. 

A  \defstyle{Routley-Meyer model}  (\defstyle{RM-model}) is a tuple 
$\its =  \langle W, O, R, ^\ast, V \rangle$, where $\langle W, O, R, ^\ast  \rangle$
is a  Routley-Meyer frame and $V: \pvar \to \powerset{W}$ is a mapping, called a \defstyle{relevant valuation}, assigning to every atomic proposition $p \in \pvar$ a set $V(p)$ of states which is \emph{upward closed}  w.r.t. $\preceq$.

\medskip 

\defstyle{Truth of a formula $A$  in an RM-model}  
$\its =  \langle W, O, R, ^\ast, V \rangle$ at a state $u \in W$, denoted   \defstyle{$\its,u\Vdash A$}, is defined as follows:

\begin{itemize}
	\item  
	$\its,u\Vdash p$ iff $u\in V(p)$;

	\item  
	$\its,u \Vdash \rt$  iff $u \in O$;

	\item  
	$\its,u \Vdash \rn A$  iff $\its,u^\ast \not\Vdash A$; 
	
	\item  
	$\its,u\Vdash A \land B$ iff $\its,u\Vdash A$ and $\its,u\Vdash B$;
	
	\item  
	$\its,u\Vdash A \lor B$ iff $\its,u\Vdash A$ or $\its,u\Vdash B$; 
	
	\item  
	$\its,u\Vdash A \ri B$ \ iff 
	for every $v,w$, if $Ruvw$ and $\its,v\Vdash A$ then $\its,w \Vdash B$. 
	
	\item  
	$\its,u\Vdash A \fu B$ \ iff 
	there exist $v,w$ such that $Rvwu$, \ $\its,v\Vdash A$ and $\its,w \Vdash B$. 
\end{itemize}

Thus, the Routley-Meyer semantics follows a standard pattern for relational semantics of modal operators. In particular, the fusion is a binary `diamond', interpreted with a ternary relation, and negation is both a unary box and diamond, interpreted via a functional binary relation.
One can show, by a routine structural induction on formulas, (cf. e.g. \cite{Routleyetal82}) that this semantics satisfies \defstyle{upward monotonicity}: for every RM-model $\its$ and a formula $A$ of \RL, the set $\extn{A}_{\its} = \{u \mid \its,u\Vdash A \} $ is upward closed.

A formula $A$ is declared \defstyle{true in an RM-model} $\its$, denoted by $\its \Vdash A$, if  $\its,o \Vdash A$  for every  $o \in O$. It is \defstyle{valid in an RM-frame $\frm$}, denoted by $\frm \Vdash A$,  iff it is true in every RM-model over that frame, and $A$ is \defstyle{RM-valid}, denoted by $ \Vdash A$,  iff it is true in every RM-model. 

All semantic notions of truth and validity defined above can be translated to FOL, resp. universal 
monadic second order, by means of a  \defstyle{standard translation}, analogous to the one applied to modal logic (cf. \cite{bent:corr2001}).
\shortver{See the details in the full paper \cite{ramics-full}.} 
\longver{The details follow in the next subsection.

\subsection{Standard translation of  \RLP to FOL} 

Clearly, Routley-Meyer frames are first-order structures for the first-order language with unary predicate symbol $O$, unary function symbol $\ast$, ternary relation symbol $R$, and individual variables $x_1, x_2, x_3, \ldots$, informally denoted $x, x', x''$ etc. We will call this language $\FOR$. Moreover, the semantics of relevance logic can be transparently expressed in $\FOR$ and every relevance formula is then equivalently translated into a formula in $\FOR$ by the following \defstyle{standard translation} $ST: \RL \to \FOR$, parametric in a first-order individual variable: 

\begin{eqnarray*}
	ST_{x}(p) &= &P(x)\\
	ST_{x}(\rt) &= &O(x)\\
	ST_{x}(\rn A) &= &\exists x'(x' = x^{\ast} \wedge \neg ST_{x'}(A))\\ 
	ST_{x}(A \land B) &= &ST_{x}(A) \land ST_{x}(B)\\
	ST_{x}(A \lor B) &= &ST_{x}(A) \lor ST_{x}(B)\\
	ST_{x}(A \fu B) &= &\exists x' x'' (Rx'x''x \wedge ST_{x'}(A) \land ST_{x''}(B))\\
	ST_{x}(A \to B) &= &\forall x'x'' (Rxx'x'' \wedge ST_{x'}(A) \to ST_{x''}(B))
\end{eqnarray*}
where $x'$ and $x''$ are fresh individual variables. 

It is routine to check that for every Routley-Meyer model \its, state $w$ in \its\ and \RL-formula $A$, it holds that $\its, w \Vdash A$ iff $\its \models ST_{x}(A)[x := w]$, where $[x := w]$ indicates that the free variable $x$ in $ ST_{x}(A)$ is interpreted as $w$.

The additional connectives of \RLP are interpreted in the same Routley-Meyer models as \RL, except that the notion of valuation need to be adjusted so that instead of $V: \pvar \to \powerset{W}$, we have $V: \atoms 
\to \powerset{W}$ and $V$ maps nominals to principal up-sets and co-nominals to complements of principal down-sets, i.e., for all $\nomi \in \nom$ and all $\cnomm \in \cnom$ we have $V(\nomi) = \ups w$ for some $w \in W$ and $V(\cnomm) = (\dwns v)^c$ for some $v \in W$.  The semantics of the additional connectives of \RLP are given as follows:

\begin{itemize}
	\item $\its, w \Vdash \nomi $  iff $w \in V(\nomi)$
	\item $\its, w \Vdash \cnomm$  iff $w \in V(\cnomm)$
	\item $\its, w \Vdash \top$  	
	\item $\its, w \not\Vdash \bot$  
	\item $\its, w \Vdash \, \rn^{\flat} \phi$  iff there is a  $v$ such that  $v^{\ast} = w$ and $\its, v \not\Vdash\phi$
	\item $\its, w \Vdash \, \rn^{\sharp} \phi$ iff for all $v$ such that $v^{\ast} = w$, it is the case that $\its, v \not\Vdash\phi$.
	\item $\its, w \Vdash \phi \coimp\psi$ iff there exists $v$  such that $v \preceq w$,  
	$\its, v \Vdash \phi$ and $\its, v \not \Vdash \psi$
	\item $\its, w \Vdash \phi \intimp \psi$ iff for all $v \succeq w$, if $\its, v \Vdash \phi$ then $\its, v \Vdash \psi$ 
	\item {$\its, w \Vdash \phi \furesfc \psi$ iff for all $v, u \in W$, if  $Rvwu$ and $\its, v \Vdash \phi$ then $\its, u \Vdash \psi$}
\end{itemize}

Under the assumption that $^\ast$ is an involution, i.e.\ that $w^{\ast\ast} = w$ for all $w \in W$, the clauses for $\rn^{\flat}$ and $\rn^{\sharp}$ become 

\begin{itemize}
	\item $\its, w \Vdash \rn^{\flat} \phi$ \ \ iff \ \ $\its, w^\ast \not\Vdash\phi$ \ \ iff \ \ $\its, w \Vdash \rn \phi$ and 
	\item $\its, w \Vdash \rn^{\sharp} \phi$ \ \ iff \ \ $\its,w^\ast \not\Vdash\phi$ \ \  iff \ \ $\its, w \Vdash \rn \phi$.\\
\end{itemize}

The standard translation $ST$ can be extended to the language \RLP. For that purpose we will add sets of individual variables $\{ y_0, y_1, y_2,\ldots\}$ and $\{ z_0, z_1, z_2,\ldots\}$ to be used for the translations of nominals and co-nominals, respectively. We extend the translation with the following clauses:

\begin{eqnarray*}
	ST_{x}(\nomj_i) &= &x \succeq y_i\\
	ST_{x}(\cnomm_i) &= &\neg (x \preceq z_i)\\
	{ST_{x}(\top)} &= & {x = x}\\
	ST_{x}(\bot) &= &\neg (x = x)\\
	ST_{x}(\rn^{\flat} \phi) &= &\exists x' ((x')^\ast = x  \wedge \neg ST_{x'}(\phi))\\
	ST_{x}(\rn^{\sharp} \phi) &= &\forall x' ((x')^\ast = x  \to \neg ST_{x'}(\phi))\\
	{ST_{x}(\phi \coimp \psi )} &= &\exists x' (x' \preceq x \wedge ST_{x'}(\phi) \wedge \neg ST_{x'}(\psi))\\
	{ST_{x}(\phi \intimp \psi)} &= &\forall x' (x' \succeq x \wedge ST_{x'}(\phi) \to ST_{x'}(\psi))\\
	{ST_{x}(\phi \furesfc \psi)} &= &\forall x' \forall x'' (Rx'xx'' \wedge ST_{x'}(\phi) \to ST_{x''}(\psi))\\
\end{eqnarray*}
where $x', x''$ are fresh individual variables, and $x \preceq x'$ is shorthand for $\exists x''(O(x'') \wedge R(x''xx'))$. 

} 

\subsection{Perfect relevant algebras and the extended language \RLP}
\label{subsec:perfectRA}

Given a Routley-Meyer frame $\frm =\langle W, R, \ast, O  \rangle$, its \defstyle{complex algebra} is the structure  \[\frm^{+}  = \langle \mathcal{P}^{\ups}(W), \cap, \cup, \complimp, \complfu, \complneg, O, W, \varnothing \rangle\] 
where  $\mathcal{P}^{\ups}(W)$ is the set of all upwards closed subsets (hereafter called \defstyle{up-sets}) of $W$, $\cap$ and $\cup$ are set-theoretic intersection and union, and for all $Y, Z \in \mathcal{P}^{\ups}(W)$ the following hold: 

\smallskip
$Y \complimp Z = \{x \in W \mid \text{ for all } y, z \in W, \text{ if } Rxyz \text{  and } y \in Y, \text{ then } z \in Z \}$, 

\smallskip
$Y \complfu Z = \{x \in W \mid  \text{ there exist } y \in Y \text{  and }  z \in Z \text{  such that } Ryzx \}$. 

\smallskip
${\complneg} Y = \{x \in W \mid x^\ast \not \in Y\}.$

It is easy to check that $\frm^{+}$ is a relevant algebra. 

An element $a$ of a lattice $\mathbb{L}$ is \defstyle{completely join-irreducible} (resp., \defstyle{completely join-prime}) if whenever $a = \bigvee S$ ($a \leq \bigvee S$) for some $S \subseteq L$, then $a = s$ ($a \leq s$) for some $s \in S$. The notions of \defstyle{meet-irreducibility} and \defstyle{primality} are defined order-dually. Complete join/meet primality implies complete join/meet irreducibility and for complete distributive lattices the notions coincide.   

A relevant algebra $\mathbb{A} = \langle A, \wedge, \vee, \fu, \ri, \rn, \rt, \top \bot \rangle$ is \defstyle{perfect} if $\langle A, \wedge, \vee, \top \bot \rangle$ is a complete, completely distributive lattice that is join-generated (resp.,
 meet-generated) by the set of its completely join-irreducible elements $J^{\infty}(\mathbb A)$ 
(resp., the set of its completely meet-irreducible elements $M^{\infty}(\mathbb A)$), while $\bigvee S \fu a = \bigvee_{s \in S} (s \fu a)$,  $a \fu \bigvee S = \bigvee_{s \in S} (a \fu s)$, $\bigvee S \ri a = \bigwedge_{s \in S} (s \ri a)$, $a \ri \bigwedge S = \bigwedge_{s \in S} (a \ri s)$, $\rn \bigvee S = \bigwedge_{s \in S} \rn s$ and $\rn \bigwedge S  = \bigvee_{s \in S} \rn s$ for all $S \subseteq A$ and $a \in A$. 
Now, in fact, every $\frm^{+}$ is a \emph{perfect} relevant algebra.
Further, every relevant algebra $\mathbb{A}$ can be compactly and densely embedded in a unique perfect relevant algebra, namely in its \emph{canonical extension} (cf. e.g. \cite{dunn2005canonical}) which we will denote $\mathbb{A}^{\delta}$. 

For any perfect distributive lattice $\mathbb A$, the map $\kappa: J^{\infty}(\mathbb A)\rightarrow M^{\infty}(\mathbb A)$ defined by $j \mapsto \bigvee\{u \in \mathbb A \mid j \not\leq u\}$ is an order isomorphism (cf.\ \cite[Sec.\ 2.3]{GNV}) when considering $J^{\infty}(\mathbb A)$ and $M^{\infty}(\mathbb A)$ as subposets of $\mathbb A$. The inverse of $\kappa$ is $\lambda: M^{\infty}(\mathbb A)\rightarrow J^{\infty}(\mathbb A)$, given by the assignment $m \mapsto \bigwedge\{u \in \mathbb A\mid u\not \leq m \}$.
From these definitions, we immediately have that, for every $u \in \mathbb A$, every $j \in J^{\infty}(\mathbb A)$ and every $m \in M^{\infty}(\mathbb A)$,
\begin{equation}\label{eq: kappa}
	j \not \leq u\ \mbox{ iff }\ u \leq \kappa(j),
\end{equation}
\begin{equation}\label{eq: lambda}
	u \not\leq m \ \mbox{ iff }\ \lambda(m)\leq u.
\end{equation}

Since in perfect relevant algebras each of $\rn$, $\vee$, $\wedge$, $\fu$ and $\ri$ preserves  or reverses arbitrary meets and/or joins in each coordinate, they are residuated in each coordinate (see e.g.\ \cite{galatos2007residuated}). The algebra therefore supports the interpretation of an extended language with connectives for the residuals of these operations. In particular, we extend the language \RL to \RLP by adding the \defstyle{left adjoint} $\rn^{\flat}$ and the  \defstyle{right adjoint} $\rn^{\sharp}$ of $\rn$ , the  \defstyle{intuitionistic (Heyting) implication} $\intimp$ (as right residual of $\wedge$), the  \defstyle{coimplication}
  $\coimp$ as the left residual of $\vee$, and the operation $\furesfc$ as the residual of $\fu$ in the second coordinate and of $\ri$ in the first coordinate.
  \longver{
  \footnote{There are different naming conventions for residuals in the literature. E.g., some authors use the terms right/left residual to refer to the residual of a binary operation in its right/left coordinate. Here we use the term \emph{right residual} for an operation which either  preserves meets or reverses joins in each of its coordinate, which implies that it is the ``right half'' of a residuated pair. Left residuals are defined order-dually. Thus, in the seven residuated pairs enumerated here for perfect relevant algebras, left residuals always appear on the left of the inequalities and right residuals on the right.}} 
  Thus, in any perfect relevant algebra $\mathbb{A}$ we have that: 

\begin{multicols}{2}
	\begin{enumerate}
		\item $\rn a \leq b$ iff $\rn^{\flat} b \leq a$
		
		\item $a \leq \rn b$  iff $b \leq \rn^{\sharp} a$
		
		\item $a \leq b \vee c$  iff $a \coimp b \leq c$
		
		\item $a \wedge b \leq c$  iff $a \leq b \intimp c$
		
		\item $a \fu b \leq c$  iff $a \leq b \ri c$
		
		\item $a \fu b \leq c$  iff $b \leq a \furesfc c$

	\end{enumerate}
\end{multicols}	

We also include in \RLP two countably infinite sets of special variables, $\nom = \{\nomj_0, \nomj_1, \nomj_2, \ldots \}$ and $\cnom = \{\cnomm_0, \cnomm_1, \cnomm_2, \ldots \}$. These are respectively called \defstyle{nominals} and \defstyle{co-nominals} and will be interpreted as ranging respectively over completely join-irreducibles and completely meet-irreducibles. Informally, we will denote nominals by $\nomi, \nomj, \nomk$, possibly with indices, while co-nominals will be denoted by $\cnomm, \cnomn$, possibly with indices. To distinguish visually from \RL, the formulas of the extended language  $\RLP$ will be denoted by lowercase greek letters, typically $\alpha, \beta, \gamma, \phi, \psi, \xi$, etc. and are defined by the following grammar: 
{\small
\begin{eqnarray*}
	\phi & =&p
	\mid \nomi
	\mid \cnomm
	\mid \top 
	\mid \bot
	\mid \rt
	\mid\, \rn \phi 
	\mid (\phi \land \phi) 
	\mid (\phi \lor \phi) 
	\mid (\phi \fu \phi) 
	\mid (\phi \ri \phi)
	\mid \, \\
	& &\rn^{\flat} \phi
	\mid \,\rn^{\sharp} \phi \mid
	(\phi \coimp \phi)
	\mid (\phi \intimp \phi)
	\mid (\phi \furesfc \phi)
\end{eqnarray*}
}
where $p \in \pvar$, $\nomi \in \nom$ and $\cnomm \in \cnom$. 
We denote 
$\atoms := \pvar \cup \nom \cup \cnom$. 
The elements of $\atoms$ will be called \defstyle{atoms}.   
An \RLP-formula is called \defstyle{pure} if it contains no propositional variables but only, possibly, nominals, co-nominals and constants. 
To each connective we assign a \defstyle{polarity type}\footnote{Also called an \defstyle{order type} (e.g.\ \cite{GNV}) or \defstyle{monotonicity type} (e.g.\ \cite{gehrke2004bounded}).} indicating whether each coordinate of its interpretation in (perfect) relevant algebras is order-preserving or order-reversing, as follows:
\begin{multicols}{2}
	\begin{enumerate}
		\item $\epsilon_{\rn} = \epsilon_{\rn^{\flat}} = \epsilon_{\rn^{\sharp}} = (-)$
		
		\item $\epsilon_{\land} = \epsilon_{\lor} = \epsilon_{\fu} = (+,+)$
		
		\item $\epsilon_{\ri} = \epsilon_{\intimp} = \epsilon_{\furesfc} = (-,+)$

		\item $\epsilon_{\coimp} = (+,-)$
	\end{enumerate}
\end{multicols}
We write $\epsilon_h(i)$ for the $i$-th coordinate of $\epsilon_h$. We now define the notions of \defstyle{positive} and \defstyle{negative occurrences} of atoms in \RLP-formulas recursively: an occurrence of an atom $a$ is positive in $a$; an occurrence of $a$ which is positive (negative) in $\phi$ is positive (negative) in $h(\psi_1, \ldots, \psi_{i-1}, \phi, \psi_{i+1}, \ldots \psi_n)$ if $\epsilon_{h}(i) = +$ and negative (positive) in $h(\psi_1, \ldots, \psi_{i-1}, \phi, \psi_{i+1}, \ldots, \psi_n)$ if $\epsilon_{h}(i) = -$.
We then say that a \defstyle{formula $\phi \in \RLP$ is positive (negative) in an atom $a$} iff all occurrences of $a$ in  $\phi$ are positive (negative). An \defstyle{inequality $\phi \leq \psi$ is positive (negative) in an atom $a$} if $\phi$ is negative (positive) in $a$ while $\psi$ is positive (negative) in $a$.

\section{Duality between perfect relevant algebras and complex algebras of Routley-Meyer frames}
\label{sec:Duality}

As already mentioned, the complex algebra $\frm^{+}$ of any Routley-Meyer frame $\frm =\langle W, R, \ast, O  \rangle$ is a perfect relevant algebra. Moreover, $J^{\infty}(\frm^{+}) = \{ \ups x \mid x \in W\}$ the set of all \defstyle{principal up-sets} $\ups x = \{ y \in W \mid y \succeq x\}$ and $M^{\infty}(\frm^{+}) = \{ (\dwns x)^c \mid x \in W \}$ the set of all set-theoretic complements of principal downwards closed subsets (hereafter called \defstyle{co-downsets}) $\dwns x = \{ y \in W \mid x \succeq y\}$. Conversely, we will show that every perfect relevant algebra is isomorphic to the complex algebra of a Routley-Meyer frame. 

\begin{lemma}\label{lem:Rn:Sharp:JIr:to:MIr}
	In a perfect relevant algebra $\mathbb A$, it is the case that ${\rn}^{\sharp}$ maps $J^{\infty}(\mathbb{A})$ into $M^{\infty}(\mathbb{A})$ and $\sim^{\flat}$ maps $M^{\infty}(\mathbb{A})$ into $J^{\infty}(\mathbb{A})$. 
\end{lemma}

\begin{proof}
\shortver{See proof in the full paper \cite{ramics-full}.}
	\longver{
	Suppose that $j \in J^{\infty}(\mathbb{A})$ and that $\bigwedge_{i \in I} a_i \leq {\rn}^{\sharp} j$. This is the case iff $j \leq {\rn} \bigwedge_{i \in I} a_i$, iff $j \leq \bigvee_{i \in I} {\rn}  a_i$. Since $j \in J^{\infty}(\mathbb{A})$, the latter is the case iff $j \leq {\rn} a_0$ for some $a_0 \in \{a_i \mid i \in I\}$, which implies $a_0 \leq {\rn}^{\sharp} j$. The argument in the case of $\sim^{\flat}$ is order-dual.
	}
\end{proof}

The following definition adapts a well-known method (see \cite{dunn2005canonical}) for obtaining dual relational structures from perfect algebras:

\begin{definition}
	The \defstyle{prime structure} of a perfect relevant algebra \\ 
	$\mathbb{A} = \langle A, \wedge, \vee, \fu, \ri, \rn, \rt, \top \bot \rangle$ is the structure $\mathbb{A}_{\bullet} = \langle J^{\infty}(\mathbb{A}), O_{\rt}, R_{\fu}, \ast_{\rn} \rangle$ where:\\
	1. $R_{\fu}abc$ iff $c \leq a \fu b$  \quad 2. $O_{\rt}\! =\! \{j \!\in\! J^{\infty}(\mathbb{A}) |  j\! \leq\! \rt \}$ \quad and \quad 3. $a^{\ast_{{\rn}}} =  \lambda(\rn^{\sharp} a)$
\end{definition}

\begin{lemma}
\label{lem2}
	$\mathbb{A}_{\bullet}$ is a Routley-Meyer frame. Moreover the order $\preceq$ on $\mathbb{A}_{\bullet}$ coincides with the dual lattice order $\geq$ restricted to $J^{\infty}(\mathbb A)$.
\end{lemma}

\begin{proof}
	We begin by noting that $b \preceq c$ iff there exists $j_{0} \in O_{\rt} = \{j \in J^{\infty}(\mathbb{A}) \mid j \leq \rt \}$ such that $R_{\fu}j_{0}bc$. By definition, the latter is equivalent to $c \leq j_{0} \fu b$ for some completely join-irreducible  $j_{0} \leq \rt$.
	By the monotonicity of $\fu$, this implies that  $c \leq \rt \fu b$   which is equivalent to $c \leq b$ by the clause 8 of Definition \ref{def:relevant:algebra}. Conversely, if $c \leq b$, then, by the same clause,  
	we have $c \leq \rt \fu b = \bigvee \{ j \in J^{\infty}(\mathbb{A}) \mid j \leq \rt \} \fu b = \bigvee \{ j \fu b \in J^{\infty}(\mathbb{A}) \mid j \leq \rt \}$. Since $c\in J^{\infty}(\mathbb{A})$, this means there is some $j_0 \in J^{\infty}(\mathbb{A})$ such that  $j \leq \rt$ and $c \leq  j \fu b$, which implies $b \preceq c$.
\longver{ 

}	
	It is clear from the construction that $\mathbb{A}_{\bullet}$ is a relevance frame. In particular, the fact that $\ast_{{\sim}}$ maps elements of $J^{\infty}(\mathbb A)$ into $J^{\infty}(\mathbb A)$ follows from the definition of $\lambda$ and Lemma \ref{lem:Rn:Sharp:JIr:to:MIr}. 
	\shortver{We verify the six defining properties of Routley-Meyer frames in 
	\cite{ramics-full}.}  
	\longver{
	We now verify the six defining properties of Routley-Meyer frames. 
	\begin{itemize}
		\item That $a \preceq a$ for all $a \in J^{\infty}(\mathbb{A})$ follows immediately from the fact that $\preceq$ coincides with $\geq$ restricted to $J^{\infty}(\mathbb{A})$.
		
		\item Properties 2, 3, and 4 follow from the definition of $R_{\fu}abc$ and the monotonicity of $\circ$ by standard arguments, see \cite{dunn2005canonical}.
		
		\item To prove property 5, suppose that $a \preceq b$. Then $b \leq a$, and so ${\rn}^{\sharp} a \leq {\rn}^{\sharp} b$, by the antitonicity of ${\rn}^{\sharp}$. Then $a^{\ast_{{\rn}}} = \lambda ({\rn}^{\sharp} a) = \bigwedge \{u \in \mathbb A \mid u \not\leq {\rn}^{\sharp} a \} \leq \bigwedge \{u \in \mathbb A \mid u \not\leq {\rn}^{\sharp} b \} = \lambda ({\rn}^{\sharp} b) = b^{\ast_{\rn}}$ and hence $b^{\ast_{\rn}} \preceq a^{\ast_{\rn}}$.
		
		\item Lastly, to verify property 6, it is clear from the definition that $O_{\rt}$ is downward closed with respect to $\leq$ restricted to $J^{\infty}(\mathbb{A})$, and hence upward closed with respect to $\preceq$.
	\end{itemize}
	}
	
\end{proof}

\begin{proposition}
\label{prop:representation}
	For any perfect relevant algebra $\mathbb{A}$ it is the case that $\mathbb{A} \simeq  (\mathbb{A}_{\bullet})^+$.
\end{proposition}

\begin{proof}
	We show that the map $\theta: \mathbb{A} \to (\mathbb{A}_{\bullet})^+$ given by $\theta(a) \mapsto \{j \in J^{\infty}(\mathbb{A}) \mid j \leq a \}$ is a relevant algebra isomorphism. 
	\shortver{See details in the full paper \cite{ramics-full}.} 
	
	\longver{
	Clearly $\theta(a)$ is a $\leq$-downset, hence a $\preceq$-upset, so $\theta(a)$ is an element of the domain of  $(\mathbb{A}_{\bullet})^+$. The injectivity of $\theta$ follows by the fact that $\mathbb{A}$ is completely join-generated by $J^{\infty}(\mathbb{A})$, while the surjectivity follows from the fact that $S = \{j \in J^{\infty}(\mathbb{A}) \mid j \leq \bigvee S\}$ for every $\leq$-downset $S \subseteq J^{\infty}(\mathbb{A})$. 
	
	Next, note that $\theta(\rt) = \{ j \in J^{\infty}(\mathbb{A}) \mid j \leq \rt \} = O_{\rt}$, $\theta(\top) =  J^{\infty}(\mathbb{A})$ and  $\theta(\bot) =  \varnothing$. 
	The facts that $\theta(a \wedge b) = \theta(a) \cap \theta (b)$ and $\theta(a \vee b) = \theta(a) \cup \theta (b)$ are immediate. 
	
	Now, we claim that $\theta(\rn a) =  \{ j \in J^{\infty}(\mathbb{A}) \mid j \leq \rn a \} = \{ j \in J^{\infty}(\mathbb{A}) \mid j^{\ast_{\rn}} \not\leq a \}$. Indeed, $j \leq \rn a$ iff $a \leq \rn^{\sharp} j$. By Lemma \ref{lem:Rn:Sharp:JIr:to:MIr}, we have $\rn^{\sharp} j \in M^{\infty}(\mathbb{A})$ so, by \eqref{eq: lambda},  $a \leq \rn^{\sharp} j$ iff  $\lambda(\rn^{\sharp} j) \not\leq a$ iff $j^{\ast_{\rn}} \not\leq a$. So $\theta(\rn a) = \{ j \in J^{\infty}(\mathbb{A}) \mid j^{\ast_{\rn}} \not\leq  a \} = \{ j \in J^{\infty}(\mathbb{A}) \mid j^{\ast_{\rn}} \not\in \theta(a) \} = \ \rn \theta(a)$.
	
	For the case of $\fu$, note that $\theta (a \fu b) = \{ j \in J^{\infty}(\mathbb{A}) \mid j \leq a \fu b \} 
	= \{ j \in J^{\infty}(\mathbb{A}) \mid j \leq \bigvee \theta(a) \fu \bigvee \theta(b) \} 
	= \{ j \in J^{\infty}(\mathbb{A}) \mid j \leq \bigvee \{ j'\fu j'' \mid j' \in \theta(a), j'' \in \theta(b)\}
	= \{ j \in J^{\infty}(\mathbb{A}) \mid (\exists j' \in \theta(a))(\exists j'' \in \theta(b))(j \leq j' \fu j'')\}
	= \{ j \in J^{\infty}(\mathbb{A}) \mid (\exists j' \in \theta(a))(\exists j'' \in \theta(b))(R_{\fu}j' j'' j)\}
	= \theta(a) \fu \theta(b)$.
	
	Lastly, for the case of $\ri$, note that  $\theta(a \ri b) = \{ j \in J^{\infty}(\mathbb{A}) \mid j \leq a \ri b\} = \{ j \in J^{\infty}(\mathbb{A}) \mid j \fu a \leq  b\}$ and $\theta(a) \to \theta(b) = \{j \in J^{\infty}(\mathbb{A}) \mid (\forall i, k \in J^{\infty}(\mathbb{A})) (R_{\fu} j i k \ \text{ and }  \ i \in \theta(a) \Rightarrow k \in \theta(b)) \}$. To establish the equality of these two sets, it is sufficient to note that:
	\begin{align*}
		&\phantom{\text{iff }} (\forall i, k \in J^{\infty}(\mathbb{A})) (R_{\fu} j i k \ \amp \ i \in \theta(a) \Rightarrow k \in \theta(b))\\
		&\text{iff } (\forall i, k \in J^{\infty}(\mathbb{A})) (k \leq j\fu i \ \amp \ i \leq a \Rightarrow k \leq b)\\
		&\text{iff } (\forall k \in J^{\infty}(\mathbb{A})) ((\exists i \in J^{\infty}(\mathbb{A}))(k \leq j\fu i \ \amp \ i \leq a) \Rightarrow k \leq b)\\
		&\text{iff } (\forall k \in J^{\infty}(\mathbb{A})) (k \leq j\fu a  \Rightarrow k \leq b)\\
		&\text{iff }  j\fu a  \leq b.\\
	\end{align*}
	}
\end{proof}

\section{The calculus of the algorithm \algo} 
\label{sec:Algo}

In this section we present a calculus of rewrite rules\footnote{These rules can be seen as instantiations of the rules of the general-purpose algorithm \alba \cite{NonDistALBA} in the context of perfect relevant algebras. However, the fact that the latter are distributive lattice expansions allows us to present simpler formulations of these rules closer to those in \cite{ALBAPaper} and, to some extent, \cite{SQEMAI}. The approximation rules presented in \cite{NonDistALBA} allow for the extraction of subformulas deep from within the consequents of quasi-inequalities, subject to certain conditions, rather than the connective-by-connective style of our presentation. Although the former style of rule is also sound in the present setting, we opted for the latter as we believe it is simpler to present since the formulation requires significantly fewer auxiliary notions.}, in the style of the algorithms SQEMA \cite{SQEMAI} and ALBA \cite{ALBAPaper,NonDistALBA}, which is sound and complete for deriving first-order frame correspondents and simultaneously proving canonicity for a large class of formulas of \RL, viz. the class of \emph{inductive (relevance) formulas} (see \cite{ConradieGoranko2021}). The algorithm \algo and its implementation, described in the next section, are based on this calculus. The algorithm accepts (inequalities of) \RLP formulas as input and, if it succeeds, it produces first-order formulas in the language of RM-frames that is valid in an RM-frame if and only if the original formulas are valid in the complex algebra of this RM-frame.

The rules manipulate \defstyle{quasi-inequalities}\footnote{In \cite{ConradieGoranko2021} these are treated set-theoretically and are called there `quasi-inclusions'.}  of \RLP formulas, i.e., expressions of the form $\phi_1 \ineq \psi_1, \ldots, \phi_n \ineq \psi_n \qi  \phi \ineq \psi$ with $\phi, \psi, \phi_i, \psi_i \in \RLP$. In the setting of relevant algebras, quasi-inequalities are considered universally quantified over all propositional variables.  Any formula $\phi \in  \RLP$ can be treated as the inequality $\rt \ineq \phi$, which is a quasi-inequality with no assumptions. The inequalities not affected by the application of the rule are regarded as a  \defstyle{context}, which will be denoted by $\Gamma$. 
Given a set of inequalities $\Gamma$, we say that $\Gamma$ is positive (negative) in an atom $a$ whenever each member of $\Gamma$ is positive (negative) in $a$. We will write $\Gamma(\alpha/p)$ for the set of inequalities obtained by uniformly substituting $\alpha$ for atom $p$ in each member of $\Gamma$. 

All rules rules that are indicated below by a double line are invertible, although the algorithm \algo only applies the approximation rules in the downward direction.

\subsubsection*{Monotone variable elimination rules:}   
{\small
\begin{prooftree}
	\def\fCenter{\ \qi \ }
	\Axiom$\Gamma(p) \fCenter \gamma(p) \ineq \beta(p)$
	\RightLabel{($\bot$)}
	\doubleLine
	\UnaryInf$\Gamma(\top/p) \fCenter \gamma(\bot/p) \ineq \beta(\bot/p)$
	
	\Axiom$\Delta(p) \fCenter \beta(p) \ineq \gamma(p)$
	\RightLabel{($\top$)}
	\doubleLine
	\UnaryInf$\Delta(\bot/p) \fCenter \beta(\top/p) \ineq \gamma(\top/p)$
	\noLine\BinaryInfC{}
\end{prooftree}
}
where $\beta(p)$ and $\Gamma$ are positive in $p$, while $\gamma(p)$ and $\Delta(p)$ are negative in $p$.

\subsubsection*{First approximation rule:} 
{\small
\begin{prooftree}
	\def\fCenter{\ \qi \ }
	\Axiom$\Gamma \fCenter \phi \ineq \psi $
	\doubleLine
	\UnaryInf$\nomj \ineq \phi, \ \psi \ineq \cnomm, \ \Gamma \fCenter \nomj \ineq \cnomm$
\end{prooftree}
}
where $\nomj$ is a nominal and $\cnomm$ is a co-nominal not occurring in the premise.

\subsubsection*{Approximation rules:} 
{\scriptsize
\begin{prooftree}
	\def\fCenter{\qi}
	\Axiom$\chi\ri \phi \ineq \cnomm, \ \Gamma \fCenter \alpha \ineq \beta$
	\RightLabel{($\ri$Appr-L)}
	\doubleLine
	\UnaryInf$\nomj\ineq\chi, \ \nomj\ri \phi \ineq \cnomm, \ \Gamma \fCenter \alpha \ineq \beta$
	
	\Axiom$\chi\ri \phi \ineq \cnomm, \ \Gamma \fCenter \alpha \ineq \beta$
	\RightLabel{($\ri$Appr-R)}
	\doubleLine
	\UnaryInf$\phi\ineq\cnomn, \ \chi \ri \cnomn \ineq \cnomm, \ \Gamma \fCenter \alpha \ineq \beta$
	\noLine\BinaryInfC{}
\end{prooftree}

\begin{prooftree}
	\def\fCenter{\qi}
	\Axiom$\nomi\ineq \chi \circ \phi, \ \Gamma \fCenter \alpha \ineq \beta$
	\RightLabel{($\circ$Appr-L)}
	\doubleLine
	\UnaryInf$\nomj\ineq\chi, \ \nomi\ineq\nomj \circ \phi, \ \Gamma \fCenter \alpha \ineq \beta$
	
	\Axiom$\nomi\ineq \chi \circ \phi, \ \Gamma \fCenter \alpha \ineq \beta$
	\RightLabel{($\circ$Appr-R)}
	\doubleLine
	\UnaryInf$\nomj \ineq \phi, \ \nomi\ineq \chi \circ \nomj, \ \Gamma \fCenter \alpha \ineq \beta$
	\noLine\BinaryInfC{}
\end{prooftree}
\begin{prooftree}
	\def\fCenter{\qi}
	\Axiom$\rn\phi\ineq \cnomm, \ \Gamma \fCenter \alpha \ineq \beta$
	\RightLabel{($\rn$Appr-L)}
	\doubleLine
	\UnaryInf$\phi\ineq\cnomn, \ \rn\cnomn\ineq\cnomm, \ \Gamma \fCenter \alpha \ineq \beta$
	
	\Axiom$\nomi\ineq \rn \phi, \ \Gamma \fCenter \alpha \ineq \beta$
	\RightLabel{($\rn$Appr-R)}
	\doubleLine
	\UnaryInf$\nomj \ineq \phi, \ \nomi\ineq \rn \nomj, \ \Gamma \fCenter \alpha \ineq \beta$
	\noLine\BinaryInfC{}
\end{prooftree}
}
where $\nomj$ a nominal and $\cnomn$ is a co-nominal not appearing in the premises. 

\subsubsection*{Residuation rules:} 
{\small
\begin{prooftree}
	\def\fCenter{\ \qi\ }
	\Axiom$\phi \ineq \chi \vee \psi, \ \Gamma \fCenter \alpha \ineq \beta$
	\RightLabel{($\vee$Res)}
	\doubleLine
	\UnaryInf$\phi \coimp\chi \ineq \psi, \ \Gamma \fCenter \alpha \ineq \beta$
	
	\Axiom$\phi \wedge \chi\ineq \psi, \ \Gamma \fCenter \alpha \ineq \beta$
	\RightLabel{($\wedge$Res)}
	\doubleLine
	\UnaryInf$ \phi \ineq \chi \intimp \psi, \ \Gamma \fCenter \alpha \ineq \beta$
	\noLine\BinaryInfC{}
\end{prooftree}
\begin{prooftree}
	\def\fCenter{\ \qi\ }
	\Axiom$\phi \ineq \chi\ri \psi, \ \Gamma \fCenter \alpha \ineq \beta$
	\RightLabel{($\to$Res)}
	\doubleLine
	\UnaryInf$\phi\circ \chi\ineq \psi, \ \Gamma \fCenter \alpha \ineq \beta$

	\Axiom$\psi \ineq \phi \furesfc \chi, \ \Gamma \fCenter \alpha \ineq \beta$
	\RightLabel{($\furesfc$Res)}
	\doubleLine
	\UnaryInf$\phi \fu \psi \ineq \chi , \ \Gamma \fCenter \alpha \ineq \beta$
	\noLine\BinaryInfC{}
\end{prooftree}
}

\subsubsection*{Adjunction rules:} 
{\small
\begin{prooftree}
	\def\fCenter{\ \qi\ }
	\Axiom$ \phi \vee \chi \ineq \psi, \ \Gamma \fCenter \alpha \ineq \beta$
	\RightLabel{($\vee$Adj)}
	\doubleLine
	\UnaryInf$ \phi \ineq \psi, \ \chi \ineq \psi, \ \Gamma \fCenter \alpha \ineq \beta$
	\Axiom$ \psi\ineq \phi \wedge \chi, \ \Gamma \fCenter \alpha \ineq \beta$
	\RightLabel{($\wedge$Adj)}
	\doubleLine
	\UnaryInf$\psi \ineq \phi, \ \psi \ineq \chi, \ \Gamma \fCenter \alpha \ineq \beta$
	\noLine\BinaryInfC{}
\end{prooftree}
\begin{prooftree}
	\def\fCenter{\ \qi\ }
	\Axiom$ \mathop{\rn} \phi \ineq \psi, \ \Gamma \fCenter \alpha \ineq \beta$
	\RightLabel{($\rn$Adj-L)}
	\doubleLine
	\UnaryInf$ \mathop{\rn^{\flat}} \psi \ineq \phi,\ \Gamma \fCenter \alpha \ineq \beta $
	\Axiom$ \phi \ineq \mathop{\rn} \psi,\ \Gamma \fCenter \alpha \ineq \beta$
	\RightLabel{($\rn$Adj-R)}
	\doubleLine
	\UnaryInf$ \psi \ineq \mathop{\rn^{\sharp}} \phi, \ \Gamma \fCenter \alpha \ineq \beta$
	\noLine\BinaryInfC{}
\end{prooftree}
}
Not to clutter the procedure with extra rules, we allow commuting the arguments of $\land$ and $\lor$ whenever needed before applying the rules ($\land$Adj) and ($\lor$Adj) above. These rules are applied exhaustively in the downward direction, and produce the same results regardless of how an expression is parenthesized.

\subsubsection*{Ackermann-rules:} 

The Right Ackermann-rule (RAR) and Left Ackermann-rule (LAR) are subject to the following conditions:
\begin{multicols}{2}
\begin{itemize}	
	\item $p$ does not occur in $\alpha$, 
	\item $\beta$ is positive in $p$,
	\item $\gamma$ is negative in $p$,
	\item $\Gamma$ is negative in $p$,
	\item $\Delta$ is positive in $p$,
\end{itemize}
\end{multicols}

{\small 
\begin{prooftree}
	\AxiomC{$\alpha \ineq p, \   \Delta(p) \qi \gamma(p) \ineq \beta(p)$}
	\RightLabel{(RAR)}
	\UnaryInfC{$\Delta(\alpha/p) \qi  \gamma(\alpha/p) \ineq \beta(\alpha/p) $}
	
	\AxiomC{$p \ineq \alpha, \ \Gamma(p) \qi \beta(p) \ineq \gamma(p)$}
	\RightLabel{(LAR)}
	\UnaryInfC{$\Gamma(\alpha/p) \qi  \beta(\alpha/p) \ineq \gamma(\alpha/p)$}
	\noLine\BinaryInfC{}
\end{prooftree}
}
Note that the rules ($\bot$) and ($\top$) are, in fact, special cases of the Ackermann-rules (RAR) and (LAR), respectively.

\subsubsection*{Simplification rules:} 

In the rules below $\Gamma$ is a possibly empty list of inequalities.
{\small
\begin{prooftree}
	\def\fCenter{\ \qi\ }
	\Axiom$\Gamma, \ \nomi \ineq \phi   \fCenter  \nomi \ineq \psi$
	\RightLabel{(Simpl-Left)}
	\doubleLine
	\UnaryInf$\Gamma \fCenter \phi \ineq \psi $
	
	\Axiom$\Gamma, \ \psi \ineq \cnomm  \fCenter  \phi \ineq \cnomm$
	\RightLabel{(Simpl-Right)}
	\doubleLine
	\UnaryInf$\Gamma  \fCenter  \phi \ineq \psi $
	\noLine\BinaryInfC{}
\end{prooftree}
}
In the rule (Simpl-Left) the nominal $\nomi$ must not occur in $\phi$, or $\psi$, or any inequality in $\Gamma$. Likewise, in the rule (Simpl-Right) the co-nominal $\cnomm$ must not occur in $\phi$, or $\psi$, or any inequality in $\Gamma$.  
These rules are usually applied in the post-processing, to eliminate nominals and  co-nominals introduced by the approximation rules.  

\shortver{ 
\begin{example}
\label{ex1}
We illustrate an application of \algo on the following formula (known as axiom B2 in \cite{Routleyetal82}):  
$(p \to q) \wedge (q \to r) \to (p \to r)$. 
In the full paper \cite{ramics-full} 
we show that the elimination phase of \algo succeeds and produces the following  pure quasi-inequality:  
	\[\nomi \circ (\nomi \circ \nomj_1)  \ineq  \cnomn_1, \   
	\nomj_1 \to \cnomn_1 \ineq \cnomm  
	\ \  \qi \ \  \nomi \ineq \cnomm.\
	\]
\end{example}
}

\longver{
\begin{example}
\label{ex1}
Here we illustrate an application of \algo on the following formula (known as axiom B2 in \cite{Routleyetal82}) also used as a running example in \cite{ConradieGoranko2021} (but, the execution presented here is optimised): 
\[(p \to q) \wedge (q \to r) \to (p \to r). \hspace{\fill} \]

\begin{enumerate}
\item 
The initial  quasi-inequality:  
	\[
	  (p \to q) \wedge (q \to r) \ineq (p \to r) 
	\]

\item Applying the First approximation rule:

	\[
	\nomi \ineq (p \to q) \wedge (q \to r), \  (p \to r) \ineq \cnomm \ \  \qi \ \  \nomi \ineq \cnomm
	\]
	
\item Applying ($\ri$Appr-Left) and ($\ri$Appr-Right) in either order produces: 	
	\[
	  \nomi  \ineq (p \to q) \wedge (q \to r), \   
	\ \nomj_1 \ineq p, \ 
	r \ineq \cnomn_1, \ \nomj_1 \to \cnomn_1 \ineq \cnomm   
	\ \  \qi \ \  \nomi \ineq \cnomm
	\]
	
\item Applying the adjunction rule ($\wedge$Adj) produces:  
	\[
	  \nomi  \ineq p \to q, \ \nomi  \ineq q \to r, \   
	\ \nomj_1 \ineq p, \ 
	r \ineq \cnomn_1, \ \nomj_1 \to \cnomn_1 \ineq \cnomm
	\ \  \qi \ \  \nomi \ineq \cnomm
	\]
	
\item Applying the Ackermann-rule with respect to $p$ produces:  
	\[
	  \nomi  \ineq  \nomj_1 \to q, \ \nomi  \ineq q \to r, \   
	\nomj_1 \to \cnomn_1 \ineq \cnomm, \  r \ineq \cnomn_1    
	\ \  \qi \ \  \nomi \ineq \cnomm
	\]

\item Applying $\circ$-residuation to $\nomi  \ineq  \nomj_1 \to q$  produces:  	

	\[
	  \nomi \circ \nomj_1 \ineq  q, \ \nomi  \ineq q \to r, \   
	\nomj_1 \to \cnomn_1 \ineq \cnomm, \  r \ineq \cnomn_1    
	\ \  \qi \ \  \nomi \ineq \cnomm
	\]

\item Applying again the Ackermann-rule now eliminates $q$: 	

	\[
	  \nomi  \ineq  (\nomi \circ \nomj_1)  \to r, \   
	\nomj_1 \to \cnomn_1 \ineq \cnomm, \  r \ineq \cnomn_1    
	\ \  \qi \ \  \nomi \ineq \cnomm
	\]

\item Applying $\circ$-residuation to $\nomi \ineq (\nomi \circ \nomj_1)  \to r$ produces:  
	\[
	  \nomi \circ (\nomi \circ \nomj_1)  \ineq  r, \   
	\nomj_1 \to \cnomn_1 \ineq \cnomm, \  r \ineq \cnomn_1    
	\ \  \qi \ \  \nomi \ineq \cnomm
	\]

\item Now, $r$ can be eliminated by one last application of the Ackermann-rule, to produce the pure  quasi-inequality:  
	\[
	  \nomi \circ (\nomi \circ \nomj_1)  \ineq  \cnomn_1, \   
	\nomj_1 \to \cnomn_1 \ineq \cnomm  
	\ \  \qi \ \  \nomi \ineq \cnomm
	\]

	Since all propositional variables have been successfully eliminated, this is the end of the  elimination phase.  
\end{enumerate}	
\end{example}
}

\section{Algorithmic description of \algo}
\label{sec:Description} 

\subsection{Pre-processing and main phase of \algo}
\label{subsec:MainPhase} 

Here we will present a deterministic algorithmic version of the procedure  \algo, which is used for the implementation.

\begin{enumerate}
	\item %
	Receive a formula $\phi$ in input.
	\item If $\phi$ is an implication $\psi\to\theta$ set $X :=\{\psi\le\theta\}$, otherwise form the initial inequality $\rt \ineq \phi$ and set $X := \{\rt \ineq \phi \}$.

	\item Now preprocess the set  $X$ by iterating steps \ref{preproc:step1}, \ref{preproc:step2} until a pass is reached in which none of the steps are applicable. 
		
	\begin{enumerate}
		
	\item\label{preproc:step1} For any $(\theta \ineq \chi) \in X$, find the first positive occurrence of $\vee$ or negative occurrence of $\wedge$ in $\theta$ which is not in the scope of any positive occurrence of $\ri$ or a negative occurrence of $\fu$. Letting $\theta(\alpha\diamond\beta)$ denote $\theta$ with the occurrence of the found subterm, where $\diamond\in\{\vee,\wedge\}$, replace $\theta \ineq \chi$ in $X$ by $\theta(\alpha)\ineq \chi,\theta(\beta)\ineq\chi$.

		\item\label{preproc:step2} For any $(\theta \ineq \chi) \in X$, find the first positive occurrence of $\wedge$ or negative occurrence of $\vee$ in $\chi$ which is not in the scope of any negative occurrence of $\ri$ or a positive occurrence of $\fu$. Again letting $\chi(\alpha\diamond\beta)$ denote $\chi$ with the found subterm, replace $\theta \ineq \chi$ in $X$ by $\theta\ineq \chi(\alpha),\theta\ineq\chi(\beta)$.
				
		The preceding two ``splitting'' steps are justified by the distributivity
		of the operations $\circ,\ri,\rn$ and the adjunction rules
		($\vee$Adj) and ($\wedge$Adj).

		\item\label{preproc:step4} Apply the monotone variable elimination rules to all inequalities in $X$ where they apply, replacing the involved inequalities in $X$ with the results. 
	\end{enumerate}

	\item Proceed separately in each inequality $\phi_i \ineq \psi_i$ in $X$. 
	Apply the first-approxima\-tion rule to $\phi_i \ineq \psi_i$ to produce the quasi-inequality
	 $\nomi \ineq \phi_i, \psi_i \ineq \cnomm \vdash \nomi \ineq \cnomm$.
	
	\item 
	As long as one of $\chi,\phi$ in the approximation rules is matched by a subformula that is neither a nominal or conominal, apply these rules exhaustively to this quasi-inequality, interleaved with the splitting steps \ref{preproc:step1}-\ref{preproc:step2}, where $X$ is the set of premises.
	The resulting quasi-inequality has premises that are irreducible with respect to
	the approximation steps and splittings. This step terminates since approximation rules are
	only applied downwards and splittings eliminate a $\land$ or $\lor$-symbol.
	
	\item
	For each variable $p$ in the quasi-inequality, and for each choice of polarity, $+p$ or $-p$, check if the right Ackermann-rule (for $+p$) or the left Ackermann-rule (for $-p$) can be applied to eliminate $p$
	from the premises of the quasi-inequality. This is done by applying the residuation and $\rn$-adjunction rules exhaustively to all premises that contain exactly one occurrence of $+p$ (or $-p$) to solve the inequality for $p$ (if possible)
	and checking that $p$ only occurs (if at all) with the opposite sign in all other premises. 
	If possible, apply the right or left Ackermann-rule. Otherwise, $p$ cannot be eliminated, in which case the next variable is tried. Backtracking is used to attempt to eliminate all variables in all possible orders and with either positive or negative polarity. If a variable cannot be eliminated in some particular quasi-inequality, then the algorithm stops and reports this failure. 
	
	\item If the elimination phase has succeeded on all quasi-inequalities, the algorithm proceeds to post-processing, including simplification and translation phases. 
	
\end{enumerate}

\subsection{Post-processing and translation to first-order logic}
\label{subsec:Post-processing} 

This phase\footnote{This is an optimised version of the post-processing procedure outlined in \cite{ConradieGoranko2021}.} applies if/when the algorithm succeeds to eliminate all variables, thus ending with \defstyle{pure quasi-inequalities}, containing only nominals and co-nominals, but no variables.  
The purpose of the post-processing is to produce a first-order condition equivalent to the pure quasi-inequality produced as a result of the main phase described in Section \ref{subsec:MainPhase}, and hence to the input formula. Each pure quasi-inequality produced in the elimination phase is post-processed separately to produce a corresponding first-order condition, and all these are then taken conjunctively to produce the corresponding first-order condition of the input formula. So, we focus on the case of a single pure  quasi-inequality. 
Computing a first-order equivalent of any pure  quasi-inequality can be done by straightforward application of the standard translation, but the result would usually be unnecessarily long and complicated. This can be compensated by additional post-translation equivalent simplifications in first-order logic, also taking into account the monotonicity conditions in Routley-Meyer frames. Instead, we have chosen to first apply some pre-translation  simplifications of the pure  quasi-inequality, using again some of the \algo rules, and then to modify the standard translation by applying it to pure inequalities, rather than to formulas, and by extending it with a number of additional clauses dependent on the type (main connective) of the formulas on both sides of these inequalities, thus applying simplifications on the fly. 
\shortver{For lack of space we have omitted the list of these additional post-processing rules, which can be found in the full paper \cite{ramics-full}.}
\longver{
We denote the modified translation function by $\tr$ and list below the additional rules that are  used for the post-processing translation phase in the implementation. In this list any newly introduced nominals $\nomi,\nomj,\nomk$ must not occur in $A,B$, and the first rule that matches a formula has priority over subsequent rules. With benign abuse of notation, we  use $\land, \lor, \to$ also to denote the classical connectives in the FO translations. 
{\small 
\begin{enumerate}
\begin{multicols}{2}
\itemsep = 2pt
\item 
$\tr(\nomi \leq \nomj)=  x_\nomj \preceq x_\nomi$

\item 
$\tr(\nomi \leq \cnomm)=  x_\nomi \not\preceq y_\cnomm$

\item 
$\tr(\nomi \leq \rt)=  Ox_\nomi$

\item 
$\tr(\nomi \leq \bot)=  \text{False}$

\item 
$\tr(\nomi \leq \top)=  \text{True}$

\item 
$\tr(\nomi \leq \rn\cnomm)=  x_\nomi^*\preceq y_\cnomm$

\item 
$\tr(\nomi \leq \rn\nomj)=  x_\nomj\not\preceq x_\nomi^*$

\item 
$\tr(\nomi \leq \rn A)= \\
 \forall x_\nomj(\tr(j\le A)\to x_\nomj\not\preceq x_\nomi^*)$

\item  
$\tr(\nomi \leq \nomj \circ \nomk)= Rx_\nomj x_\nomk x_\nomi$

\item 
$\tr(\nomi \leq \nomj \circ B)= \\ 
\exists x_\nomk (\tr(\nomk \leq B)\land Rx_\nomj x_\nomk x_\nomi)$

\item 
$\tr(\nomi \leq A \circ B)= \\
 \exists x_\nomj(\tr(\nomj \leq A) \land \tr(\nomi\leq \nomj {\circ} B))$

\item
$\tr(\nomi \le A\to B)= \tr(\nomi\circ A\le B)$ 

\item $\tr(\nomi \le A\furesfc B)= \tr(A\circ\nomi\le B)$

\item
$\tr(\nomi \le A\Rightarrow B)= \tr(\nomi\land A\le B)$

\item
$\tr(\nomi\le A\!\land\! B)\!=\!\tr(\nomi\le A) \land \tr(\nomi\le B)$

\item
$\tr(\nomi\le A\!\lor\! B)\!=\! \tr(\nomi\le A) \lor \tr(\nomi\le B)$

\item 
$\tr(\cnomn \leq \cnomm)=  y_\cnomm \preceq y_\cnomn$

\item 
$\tr(\rt \leq \cnomm)=  \neg Oy_\cnomm$

\item 
$\tr(\bot\le\cnomm)=  \text{True}$

\item 
$\tr(\top\le\cnomm)=  \text{False}$

\item 
$\tr(\rn\cnomn \leq \cnomm)=  y_\cnomm^* \not\preceq y_\cnomn$

\item 
$\tr(\rn\nomj \leq \cnomm)=  x_\nomj\preceq y_\cnomm^*$

\item 
$\tr(\rn A \leq \cnomm)= \\ \exists x_\nomj(\tr(j\le A)\land x_\nomj\preceq y_\cnomm^*)$

\item 
$\tr(\nomi \circ \nomj \leq \cnomm)= \lnot R x_\nomi x_\nomj y_\cnomm$

\item  
$\tr(\nomi \circ B \leq \cnomm)= \\ 
 \forall x_\nomj (\tr(\nomj \leq B) \to  \lnot Rx_\nomi x_\nomj y_\cnomm)$

\item  
$\tr(A \circ B \leq \cnomm)= \\ 
\forall x_\nomi (\tr(\nomi \leq A)\land\tr(\nomi\circ B\leq\cnomm)$

\item  
$\tr(A \Rightarrow B \leq \cnomm)= \\
\forall x_\nomi (\tr(\nomi \leq A\Rightarrow B) \to \tr(\nomi\le\cnomm))$ 

\item
$\tr(A\coimp B\le\cnomm)= \tr(A\le B\lor\cnomm)$

\item
$\tr(A\land B\le\cnomm)= \\ 
\tr(A\le \cnomm)\lor \tr(B\le\cnomm)$

\item
$\tr(A\lor B\le\cnomm)= \\ 
\tr(A\le \cnomm)\land \tr(B\le\cnomm)$

\item 
$\tr(A \leq B)= \\ 
 \forall x_\nomj (\tr(\nomj \leq A) \to  \tr(\nomj \leq B))$
\end{multicols}
\end{enumerate}
}
We note that the translation $\tr$ is not restricted to pure quasi-inequalities and can be applied to arbitrary pure formulas.
}

\shortver{
The resulting  modified translation $\tr$ is not restricted to pure quasi-inequalities and can be applied to arbitrary pure formulas.

The post-processing of the pure  quasi-inequality produced in  Example \ref{ex1} 	
using the translation $\tr$ is illustrated in the  full paper \cite{ramics-full}. 
The resulting first-order formula is 
$\forall x_\nomi, x_{\nomj}, x_{\nomj_1}, y_{\cnomn_1} (Rx_\nomi x_{\nomj_1}y_{\cnomn_1}\to \exists x_\nomj (Rx_\nomi x_{\nomj_1} x_\nomj  \land R x_\nomi x_\nomj y_{\cnomn_1}))$ which is equivalent to the first-order condition known from \cite{Routleyetal82} for the axiom B2, and to the one computed by the implementation of \algo reported here. 
}

\longver{
\begin{example}
\label{ex2}
The post-processing using the modified translation is illustrated on the  following pure  quasi-inequality produced in  Example \ref{ex1}: 	
\[
	  \nomi \circ (\nomi \circ \nomj_1)  \ineq  \cnomn_1, \   
	\nomj_1 \to \cnomn_1 \ineq \cnomm  
	\ \  \qi \ \  \nomi \ineq \cnomm
	\]

\begin{enumerate}

	\item Applying the  right Simplification rule (Simpl-Right) produces: 
\[
	  \nomi \circ (\nomi \circ \nomj_1)  \ineq  \cnomn_1, \   
	\ \  \qi \ \  \nomi \ineq \nomj_1 \to \cnomn_1
	\]

	\item Now, applying the $\tr$ translation steps above produces: 
	
$\tr(\nomi \circ (\nomi \circ \nomj_1)\leq \cnomn_1) \implies \tr(\nomi \ineq \nomj_1 \to \cnomn_1)$

$= 
\forall x_\nomj (\tr(\nomj \leq \nomi\circ\nomj_1)  \implies   \lnot R x_\nomi x_\nomj y_{\cnomn_1})
\implies \tr(\nomi \circ \nomj_1\leq \cnomn_1)$ \ \  (rule 25, rule 12)

$= \forall x_\nomj (Rx_\nomi x_{\nomj_1} x_\nomj  \implies   \lnot R x_\nomi x_\nomj y_{\cnomn_1})\implies \lnot Rx_\nomi x_{\nomj_1}y_{\cnomn_1}$ \ \  (rule 9, rule 24)

$=Rx_\nomi x_{\nomj_1}y_{\cnomn_1}\implies \exists x_\nomj (Rx_\nomi x_{\nomj_1} x_\nomj  \land R x_\nomi x_\nomj y_{\cnomn_1})$ \ \  (contraposition)

 Up to variable renaming, this is equivalent to the first-order condition known from \cite{Routleyetal82} for the axiom B2, and to the one computed by the implementation of \algo reported here. 

\end{enumerate}
\end{example}
}

\section{Implementation of \algo}
\label{sec:implementation} 

\longver{\subsection{Description of the \algo  implementation}} 

Here we give a brief description of an implementation of \algo  in Python, based on the description given in Section \ref{sec:Description}. The input is a \LaTeX\ string using the standard syntax of relevance logic expressions. Intuitionistic implication $\Rightarrow$, coimplication $\coimp$, the right residual $\furesfc$ of $\circ$, 
and the adjoints $\sim^\sharp$ and $\sim^\flat$ can also appear in an input formula. The expression is parsed with a simple top-down Pratt parser \cite{DBLP:conf/popl/Pratt73} using standard rules of precedence. For well-formed formulas, an abstract syntax tree (AST) based on Python dictionaries and lists of arguments is created for each formula.
For example, the formula $A=$ ``$p\to q{\land}\mathbf t$'' is translated to the internal representation

\verb| A={"id":"\to","a":[ |

\verb|    {"id":"p","a":[]},|

\verb|    {"id":"\land","a":[{"id":"q","a":[]},{"id":"\mathbf t","a":[]}]}|

\verb|  ]}|.

The implication symbol $\to$ is referenced by \verb|A.id| and the two arguments are \verb|A.a[0]| and \verb|A.a[1]|.

Five short recursive Python functions are used to transform the AST representation step-by-step according to the specific groups of \algo  transformation rules. The function \verb|preprocess(st)| takes a \LaTeX\ string \verb|st| as input and parses it to an AST which we refer to as \verb|A|. If the formula \verb|A| is not well-formed, an error-string is returned. If it has a top-level $\to$ symbol, it is replaced with a $\le$ to turn the formula into an inequality, 
and otherwise the equivalent inequality $\mathbf t\le \mathtt A$ is constructed. 
Subsequently the splitting rules and monotonicity rules from Section~\ref{sec:Algo} are applied and the resulting list of inequalities is returned.

For example, with \verb|r"p\to q\land\mathbf t"| as input, the formula is parsed, rewritten as $p\le q{\land}\mathbf t$, then the splitting rules produce the list $[p\le q,p\le\mathbf t]$ and monotonicity returns $[\top\le\bot,\top\le\mathbf t]$.

The function \verb|approximate(As)| takes this list as input, and applies the first approximation rule to each formula, followed by all possible left and right approximations interleaved with further applications of the splitting rule. The result is a list of quasi-equations that always have conclusion $\mathbf i\le \mathbf m$ and premises that are irreducible with respect to the
approximation and splitting rules.

The function \verb|eliminate(As)| then attempts to apply the Ackermann-rules to each quasi-equations by selecting each variable, first with positive polarity and, if that does not succeed, then with negative polarity. Backtracking is used to ensure that all variables
are tried in all possible orders. If for some quasi-equations none of the variable orders allow all variables to be eliminated, then the function reports this result. On the other hand, if for each quasi-equations some variable order succeeds to eliminate all formula variables
then the resulting list of pure quasi-equations (i.e., containing no formula variables, but only nominals or co-nominals) is returned.

Since these pure quasi-equations contain redundant premises, the function \verb|simplify(As)| is used to eliminate them, and to also apply the left and right simplification rules.
Finally the variant of the standard translation described in Section \ref{subsec:Post-processing} 
is applied to the pure quasi-equations and produces a first-order formula on the Routley-Meyer frames. 

The Python code can be used in any Jupyter notebook, with the output
displayed in standard mathematical notation. No special installation is needed to use
the program in a personal Jupyter notebook or in a public cloud-based notebook such as 
Colab.google.com, and the output can be pasted into standard \LaTeX\ documents. Moreover 
the program can be easily extended to handle the syntax of
other suitable logics and lattice-ordered algebras. The resulting formula can also be translated to TPTP, Prover9 or SPASS syntax.
The Python code is available at \url{github.com/jipsen/PEARL} in the form of a Jupyter notebook. It can also be copied and used directly in a browser at \url{https://colab.research.google.com/drive/1p0PTkmyq7vTWgYDxCTFHVRwjaLeT45uX?usp=sharing}.
\shortver{
In the full paper \cite{ramics-full} we provide some examples of output from the \algo  implementation. 
}

\subsection{Two examples of output from the \algo  implementation} 

\begin{itemize}

\item
Input command: \verb|pearl(|$(A\to B)\land (B\to C)\to (A\to C)$\verb|, "latex")|

\item
Translate to (list of) initial inequalit(ies): $[(A\to B)\land (B\to C)\le A\to C]$

\item
Approximation phase: \\
 $\mathbf i\le A\to B,\ \mathbf i\le B\to C,\ \mathbf j_1\to \mathbf n_1\le \mathbf m,\ C\le \mathbf n_1,\ \mathbf j_1\le A\quad\implies\quad \mathbf i\le \mathbf m$

\item
Order of variables during the elimination phase: $['+A', '+C', '+B']$

\item
Elimination phase: $\mathbf j_1\to \mathbf n_1\le \mathbf m,\ \mathbf i\circ (\mathbf i\circ \mathbf j_1)\le \mathbf n_1\quad\implies\quad \mathbf i\le \mathbf m$

\item
Apply simplification rules: $\mathbf i\circ (\mathbf i\circ \mathbf j_1)\le \mathbf n_1\quad\implies\quad \mathbf i\le \mathbf j_1\to \mathbf n_1$

\item
Apply $\tr$ rules: $\forall x_2 (R x_0x_1x_2\implies \neg (R x_0x_2y_1))\implies \neg (R x_0x_1y_1)$

\item
Contrapose and simplify: $Rx_0x_1y_1\implies \exists x_2 (Rx_0 x_1 x_2  \land R x_0 x_2 y_1)$
\end{itemize}

\begin{itemize}
\item
Input command: \verb|pearl(|$A\to ({\sim} A\to B)$\verb|, "latex")|

\item
Initial inequality after monotone variable elimination: $[A\le {\sim} A\to \bot]$

\item
Approximation phase: \\ 
$\mathbf i\le A,\ \mathbf j_1\to \mathbf n_1\le \mathbf m,\ \bot\le \mathbf n_1,\ \mathbf j_1\le {\sim} 
\mathbf n_2,\ A\le \mathbf n_2\quad\implies\quad \mathbf i\le \mathbf m$

\item
Elimination phase: $\mathbf j_1\to \mathbf n_1\le \mathbf m,\ \bot\le \mathbf n_1,\ \mathbf j_1\le {\sim} \mathbf n_2,\ 
\mathbf i\le \mathbf n_2\quad\implies\quad \mathbf i\le \mathbf m$

\item
Apply simplification rules: $\mathbf j_1\le {\sim} \mathbf n_2\quad\implies\quad \mathbf n_2\le \mathbf j_1\to \mathbf n_1$

\item
Apply $\tr$ rules: $x_1^*\preceq y_2\implies \forall x_2 (R x_2x_1y_1\implies x_2\preceq y_2)$
\end{itemize}

\section{Canonicity and applications to BI-logic and relation algebras}
\label{sec:results} 

Here we report on some new theoretical and practical results related to the theory and implementation of \algo. 
We begin with a theoretical result, which, for lack of space, we only sketch here.  

\begin{theorem} 
\label{thm:canonicity} 
The validity of all \RLP-formulas on which \algo succeeds is preserved under canonical extensions of relevant algebras.
\end{theorem} 
\begin{proof}
	Let $\phi \leq \psi$ be an \RL-inequality on which \algo succeeds and let $\mathbb{A}$ be a relevant algebra. Let $\algo(\phi \leq \psi)$ denote the purified quasi-inequality produced from input $\phi \leq \psi$. For any \RLP quasi-inequality $\Gamma \implies \alpha \leq \beta$, we write $\mathbb{A}^{\delta} \models_{\mathbb{A}} \Gamma \implies \alpha \leq \beta$ to indicate that $\Gamma \implies \alpha \leq \beta$ is true in $\mathbb{A}^{\delta}$ under all assignments that send propositional variables to elements of the original algebra $\mathbb{A}$ (and nominals to $J^{\infty}(\mathbb{A})$ and co-nominals to $M^{\infty}(\mathbb{A})$) while, as usual, $\mathbb{A}^{\delta} \models \Gamma \implies \alpha \leq \beta$ indicates truth under \emph{all} assignments. The following chain of equivalences establishes the canonicity of $\phi \leq \psi$:

	\begin{center}
		\begin{tabular}{l c l}
			$\mathbb{A} \ \ \models \phi \leq \psi$ & &$\mathbb{A}^{\delta} \models \phi \leq \psi$\\
			$\ \ \ \ \ \ \ \ \ \ \ \Updownarrow$ \\
			$\mathbb{A}^{\delta} \models_{\mathbb{A}} \phi \leq \psi$ & & \ \ \ \ \ \ \ \ \ \ \  $\Updownarrow $\\
			$\ \ \ \ \ \ \ \ \ \ \ \Updownarrow$\\
			$\mathbb{A}^{\delta} \models_{\mathbb{A}} \algo(\phi \leq \psi)$
			&\ \ \ $\Leftrightarrow$ \ \ \ &$\mathbb{A}^{\delta} \models \algo(\phi \leq \psi)$
		\end{tabular}
	\end{center}
	The uppermost bi-implication on the left is immediate by the way we defined $\models_{\mathbb{A}}$ and the fact that $\mathbb{A}$ is a subalgebra of $\mathbb{A}^{\delta}$. The lower bi-implication on the left follows by that fact that, if a quasi-inequality $\Delta' \implies \gamma' \leq \chi'$ is obtained from another, $\Delta \implies \gamma \leq \chi$, through the application of \algo rules, then $\mathbb{A}^{\delta} \models_{\mathbb{A}} \Delta \implies \gamma \leq \chi$ iff $\mathbb{A}^{\delta} \models_{\mathbb{A}} \Delta' \implies \gamma' \leq \chi'$. This is straightforward to check for all rules except the Ackermann-rules. We refer the reader to \cite{ALBAPaper} and/or \cite{NonDistALBA} for the details of the latter.  The horizontal bi-implication follows from the facts that, by assumption, $\algo(\phi \leq \psi)$ is pure, and that restricting assignments of propositional variables to elements of $\mathbb{A}$ is vacuous for pure formulas, as they contain no propositional variables. The bi-implication on the right follows  by the soundness of all \algo rules on perfect algebras, which is routine to verity. 
\end{proof}

Via the discrete duality between perfect relevant algebras and Routley-Meyer frames established in 
Section \ref{sec:Duality}, it follows that all \RLP-formulas on which \algo succeeds axiomatise logics which are complete with respect to their respective first-order definable classes of Routley-Meyer frames.

As mentioned in the introduction, a large syntactically defined class of \emph{inductive relevance formulas} in \RL is defined in \cite{ConradieGoranko2021}, where it is shown that \algo succeeds for all such formulas and correctly computes their equivalent with respect to frame validity first-order definable conditions on Routley-Meyer frames. Therefore, all inductive \RLP-formulas are canonical. This result generalizes the ``canonicity via correspondence" result in \cite{DBLP:journals/sLogica/Urquhart96}, applied there to the fragment of \RL involving of all specific relevance logic connectives only the fusion.  

We can now state the results above applied to the specific  implementation of \algo reported here. However, the proof of the correctness of the implementation is beyond the scope of this paper. 
Still, we can report that the implementation has succeeded on all axioms A1-A9, B1-B30, and D1-D8 listed in the appendix of \cite{ConradieGoranko2021}, copied there from \cite{Routleyetal82}, and has computed first-order conditions equivalent to those known from the literature. 

\medskip

Bunched implication logic  \cite{pym2002semantics} is closely related to a negation-free relevance logic. The algebraic
semantics of bunched implication logic is given by bunched implication algebras, or BI-algebras.
They are defined by axioms \textit{1-3} and \textit{7-9} of Definition~\ref{def:relevant:algebra} together with a new binary operation symbol $\Rightarrow$ such that
	\begin{enumerate}
 \begin{multicols}{2}
		\item[\textit{10.}] $a \wedge b \leq c$ iff $a \leq b \Rightarrow c$  
		{\small \ (hence $\Rightarrow$ \\ is a Heyting algebra implication)} 
		\item[\textit{11.}] $(a \fu b) \fu c = a \fu (b \fu c)$,  
		\item[\textit{12.}] $a \fu b = b \fu a$.
\end{multicols}		
	\end{enumerate}
The steps of the \algo  algorithm are not affected by these addition axioms (although additional rules for the associativity and commutativity of $\fu$ could be added), and the relational semantic structures of BI-logic and BI-algebras are precisely Routley-Meyer frames.
However in BI-logic the notation differs slightly, since $\ri, \fu, \Rightarrow$ are replaced by $-\!*,*,\to$, and this alternative notation is user-selectable in the implementation. 

Lastly, we note that the algorithm \algo can also be applied to relation algebras, as  they form a subvariety of relevant algebras extended with a Heyting implication $\Rightarrow$. 
An axiomatization of relation algebras in this setting consists of axioms of relevant algebras (\textit{1-9} from Definition~\ref{def:relevant:algebra}), \textit{10, 11} above and\footnote{While this equational basis for relation algebras appears to be quite long, it can be shown that axioms \textit{3-7} are redundant. Hence,  it is comparable in length to the original axiomatization of relation algebras.}  
	\begin{enumerate}
	 \begin{multicols}{2}
		\item[\textit{13.}] $(x\Rightarrow\bot)\Rightarrow\bot=x$ \\ 
			{\small (hence $\Rightarrow$ is a  
		classical implication \\ and $x\Rightarrow \bot$ is denoted $\neg x$),} 
		\item[\textit{14.}] $x\ri y=\rn(\rn y\fu x)$,
		\item[\textit{15.}] $x^\smallsmile=\rn(x\Rightarrow\bot)$,  
		\item[\textit{16.}] $(x\fu y)^\smallsmile=y^\smallsmile\fu x^\smallsmile$.
	\end{multicols}	
	\end{enumerate}
Axiom \textit{13} ensures that the lattice structure is a Boolean algebra, hence the
partial order in the Routley-Meyer frames of a relation algebra is an antichain. 
In the theory of relation algebras these frames are known as `atom structures', defined in \cite[Def. 2.1]{Maddux1982}. For the application of \algo  to relation algebras, it suffices to replace
the converse operation by the term $\rn(x\Rightarrow\bot)$ and to interpret any $\preceq$ symbol in the resulting first-order formula as an equality symbol. Note that relevant negation $\rn x$ can, in turn, also be defined via the relation algebra term $(\neg x)^\smallsmile$. While there is a long history of Sahlqvist formulas and correspondence theory for Boolean algebras with operators \cite{deRijVen1995, jonsson1994canonicity}, it is interesting to note that the PEARL algorithm and its implementation can be adapted to relation algebras and covers the more general class of inductive formulas.

\longver{
\section{Concluding remarks} 
\label{sec:Concluding} 
In this paper we have reintroduced the algorithm \algo from \cite{ConradieGoranko2021} as an algorithm which manipulates quasi-inequalities interpreted over perfect relevant algebras. Purely in these terms, \algo is an instantiation in the setting of relevant algebras of the general theory developed in \cite{NonDistALBA}. While this general theory prescribes a set of rules which are sufficient to produce an equivalent pure quasi-inequality out of any inductive inequality, more was required to produce an efficiently implementable algorithm producing reasonably optimal in size versions of first-order correspondents of the various \RL axioms on Routley-Meyer frames out of these pure quasi inequalities. In particular, detailed algorithmic specifications and strategic choices in the pre-processing, main, and post-processing phases (Sections \ref{subsec:MainPhase} and \ref{subsec:Post-processing}) were required and a specialized post-processing and translation procedure, refining the normal standard translation, is developed in Section \ref{subsec:Post-processing}. 
 
Many questions in the correspondence theory of relevance logic and the related algebraic structures considered here remain unanswered. We mention here just three directions for future research. Firstly, the current results can also be expanded to deal with relevant modal logic (see e.g.  \cite{seki2003sahlqvist}), and this should be reasonably straightforward. Secondly, while we now have an implemented algorithm for finding first-order frame correspondence for a wide class of \RL-formulas, a theory of inverse correspondence (like that of Kracht for modal logic \cite{kracht1993completeness,Kracht:Tools:techniques}) which would find \RL-formulas defining given first-order properties of RM-frames is still to be developed. Thirdly, relativized correspondence phenomena like those found in modal logic (see e.g. \cite{van1976modal, Balbiani:Georgiev:Tinchev:2017}) remain to be systematically investigated.

Last, but not least, we hope and expect that the present work, in particular the implementation of the algorithm \algo, will find many useful applications aiding the further research on relevance logics, bunched implication logics, and relation algebras. 
}

\shortver{
\section{Concluding remarks}\label{sec:Concluding}%
In this paper we have re-interpreted the algorithm \algo from \cite{ConradieGoranko2021} as an algorithm which manipulates quasi-inequalities interpreted over perfect relevant algebras. Implementing the algorithm in a way that produces reasonably optimal (in size) versions of first-order correspondents required detailed specifications and strategic choices in the pre-processing, main, and post-processing phases (Sections \ref{subsec:MainPhase} and \ref{subsec:Post-processing}) and in the specialized post-processing and translation procedure, refining the normal standard translation, developed in Section \ref{subsec:Post-processing}. 
It is easy to see that the complexity of the problem solved by \algo is in NP-time because, once  the correct ordering or elimination of the variables is selected, \algo completes its work in polynomial time. However, theoretically, it may take trying an exponential number of such orderings until success. Whether this is possible is not yet known, so the optimal complexity of the problem is still under investigation. 
}

%
%
%

\bibliographystyle{splncs04}
\bibliography{PEARL-short}

\end{document}